\documentclass[pra,twocolumn,preprintnumbers,amsmath,amssymb,showpacs,nofootinbib,floatfix]{revtex4}

\usepackage{graphicx,bm}

\makeatletter
\def\graphicscale{\twocolumn@sw{0.3}{0.4}}
\def\graphicthreescale{\twocolumn@sw{0.3}{0.4}}

\begin{document}

\title{Critical behavior at the spatial boundary of \\

a trapped inhomogeneous Bose-Einstein condensate}

\author{Francesco Delfino and Ettore Vicari}

\address{Dipartimento di Fisica dell'Universit\`a di Pisa and
  INFN, Largo Pontecorvo 3, I-56127 Pisa, Italy} 

\date{\today}

\begin{abstract}

We investigate some aspects of the Bose-Einstein condensation (BEC) of
quantum gases in the presence of inhomogeneous conditions. We consider
three-dimensional (3D) quantum gases trapped by an external potential
when the temperature is sufficiently low to show a BEC phase region
around the center of the trap.  If the trap is sufficiently large,
different phases may coexist in different space regions, when moving
from the center of the trap.  We show that the quantum gas develops a
peculiar critical behavior at the boundary of the BEC region, whose
scaling behavior is controlled by the universality class of the
homogenous BEC transition.  We provide numerical evidence of this
phenomenon, for lattice atomic gases modeled by the 3D Bose-Hubbard
Hamiltonian.

\end{abstract}


\maketitle



\section{Introduction}
\label{intro}

Three-dimensional (3D) bosonic gases show the phenomenon of the
Bose-Einstein condensation (BEC), which gives rise to a
finite-temperature phase transition separating the high-temperature
normal phase from the low-temperature superfluid BEC phase.  The
formation and properties of the BEC in quantum gases have been
investigated by several experiments with cold atoms in harmonic traps,
see, e.g.,
Refs.~\cite{Andrews-etal-97,Stenger-etal-99,Hagley-etal-99,BHE-00,
  Cacciapuoti-etal-03,Hellweg-etal-03,Ritter-etal-07,BDZ-08,DGPS-99,
  PSW-01,Dettmer-etal-01,Hellweg-etal-02}.  For generic 3D traps the
coherence length turns out to be of the same size of the condensate (a
substantial phase decoherence is only expected for a very elongated
BEC, along the longer
direction~\cite{PSW-01,Dettmer-etal-01,Hellweg-etal-02,Mathey-etal-10,
  GCP-12,CDMV-16}).  The critical properties of 3D quantum gases at
their BEC phase transition, i.e. when the condensate begins forming,
have been investigated theoretically and experimentally, see, e.g.,
Refs.~\cite{DRBOKE-07,DZZH-07,BB-09,CV-09,ZKKT-09,
  Trotzky-etal-10,HZ-10,PPS-10,NNCS-10,ZKKT-10,QSS-10,FCMCW-11,HM-11,
  Pollet-12,CR-12,CTV-13,CN-14,CNPV-15,CNPV-16}.  In particular, the
inhomogeneous conditions due to the spatially-dependent trapping
potential give rise to a universal distortion of the critical behavior
of the corresponding homogenous systems, i.e., without the external
spatially-dependent trapping potential.

In this paper we consider 3D bosonic gases trapped by an external
harmonic potential, when the temperature is sufficiently low to show a
BEC-phase region around the center of the trap.  Due to the trapping
potential, the BEC region is generally spatially limited.  Therefore,
when moving from the center of the trap, the quantum gas passes from
the BEC phase around the center of the trap (where space coherence is
essentially described by spin waves) to a normal phase far from the
center.  We study the behavior of the correlation functions around the
center of the trap and at the boundary of the BEC.  We point out that
the quantum many-body system develops a peculiar critical behavior at
the boundary of the BEC region, with a nontrivial scaling behavior
controlled by the universality class of the homogenous BEC transition.
We provide some numerical evidence of this phenomenon in lattice
atomic gases modeled by the Bose-Hubbard (BH)
Hamiltonian~\cite{FWGF-89,JBCGZ-98}.

The paper is organized as follows.  Sec.~\ref{stbh} presents the 3D BH
model, and the main features of its finite-temperature phase diagram.
In Sec.~\ref{critspabec} we discuss the effects of the inhomogeneous
spatial conditions when the temperature is sufficiently low to show a
BEC phase around the center of the trap; the scaling behaviors at the
spatial boundary of the BEC are put forward. Sec.~\ref{numres}
presents some numerical results for the 3D BH model, supporting the
scaling ansatz at the spatial region between the BEC and normal-phase
regions.  Finally in Sec.~\ref{conclu} we draw our conclusions.  The
appendix reports the calculation of the critical exponent $\theta$ of
the scaling theory describing the behavior of the one-particle
correlation function at the boundary of the BEC.

\section{The 3D Bose-Hubbard  model}
\label{stbh}

The 3D BH model~\cite{FWGF-89} is a realistic model of bosonic atoms
in optical lattices~\cite{JBCGZ-98}.  Its Hamiltonian reads
\begin{eqnarray}
H &=& - t \sum_{\langle {\bf x}{\bf y}\rangle} (b_{\bf y}^\dagger b_{\bf x}+
b_{\bf x}^\dagger b_{\bf y}) + \label{bhm}\\
&+&{U\over 2} \sum_{\bf x} n_{\bf x}(n_{\bf x}-1) - \mu \sum_{\bf x} n_{\bf x}\,,
\nonumber
\end{eqnarray}
where $b_{\bf x}$ is a bosonic operator, $n_{\bf x}\equiv b_{\bf
  x}^\dagger b_{\bf x}$ is the particle density operator, the sums run
over the bonds ${\langle {\bf x} {\bf y} \rangle }$ and the sites
${\bf x}$ of a cubic lattice, $a=1$ is the lattice spacing.  The
one-particle correlation function
\begin{eqnarray}
G({\bf x},{\bf y};T) \equiv \langle b_{\bf x}^\dagger b_{\bf y}
\rangle \equiv  {{\rm Tr} \;b_{\bf x}^\dagger b_{\bf y} \, e^{-H/T}\over
{\rm Tr} \;e^{-H/T}}
\label{gbdef}
\end{eqnarray}
provides information on the phase coherence properties of the BEC.

\begin{figure}
\includegraphics{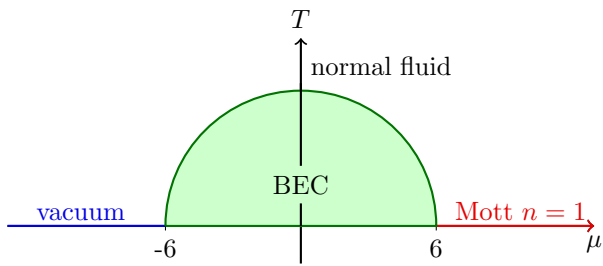}
\caption{Sketch of the $T$-$\mu$ (in unit of the hopping parameter
  $t$) phase diagram of the 3D BH model in the hard-core $U\to\infty$
  limit.  The BEC phase is restricted to the region between $\mu=-6$
  and $\mu=6$.  It is bounded by a BEC transition line $T_c(\mu)$,
  which satisfies $T_c(\mu)=T_c(-\mu)$ due to the particle-hole
  symmetry of the hard-core BH model.  Its maximum occurs at $\mu=0$,
  where~\cite{CN-14} $T_c(\mu=0)= 2.0160(1)$.  We also know
  that~\cite{CTV-13} $T_c(\mu=\pm 4)=1.4820(2)$. At $T=0$ two further
  quantum phases exist: the vacuum phase ($\mu<-6$) and the
  incompressible $n=1$ Mott phase ($\mu>6$).  }
\label{3dphasedia}
\end{figure}

The phase diagram of the 3D BH model and its critical behaviors have
been much investigated, see
e.g. Refs.~\cite{FWGF-89,CPS-07,CR-12,CTV-13,CN-14,CNPV-15}.  Their
$T$-$\mu$ phase diagram presents a finite-temperature BEC transition
line related to the formation of the condensate, i.e. the accumulation
of a macroscopic number of atoms in a single quantum state.  In
Fig.~\ref{3dphasedia} we sketch the phase diagram in the hard-core
$U\to\infty$ limit.  The BEC phase extends below the BEC transition
line $T_c(\mu)$.

The condensate wave function provides the complex order parameter of
the BEC transition. Thus its critical behavior belongs to the 3D XY
universality class characterized by the spontaneous breaking of an
Abelian U(1) symmetry~\cite{PV-02}.  In particular, around the
transition point $T_c$, the one-particle correlation function of
infinite homogenous systems behaves as
\begin{eqnarray}
&&G({\bf x}_1,{\bf x}_2;T) \approx  
\xi^{-1-\eta} {\cal G}(|{\bf x_1}-{\bf x}_2|/\xi),
\label{hocrbeh}\\
&&G({\bf x}_1,{\bf x}_2;T=T_c) \sim 
{1\over |{\bf x}_1-{\bf x}_2|^{1+\eta}},
\label{hocrbehtc}
\end{eqnarray}
where $\xi \sim \tau^{-\nu}$ is the diverging correlation length of
the critical condensing modes, and $\tau\equiv T/T_c-1$ is the reduced
temperature. The critical exponents $\nu$ and $\eta$ are those
associated with the 3D XY universality class~\cite{CHPV-06},
i.e., $\nu=0.6717(1)$ and $\eta=0.0381(2)$.

A common feature of the experiments with cold atoms~\cite{BDZ-08} is
the presence of an external potential $V({\bf x})$ coupled to the
particle density, which traps the particles within a limited space
region.  In the experiments $V({\bf x})$ is usually effectively
harmonic.  We consider a rotationally-invariant harmonic potential
(most results can be straightforwardly extended to more general cases)
\begin{eqnarray}
V(r)= v^2 r^2,\qquad r\equiv |{\bf x}|,
\label{potential}
\end{eqnarray}
where $r$ is the distance from the center of the trap, which we locate
at the origin ${\bf x}=0$.  This trapping force is taken into account
by adding a further term to the BH Hamiltonian (\ref{bhm}), i.e.,
\begin{eqnarray}
H_t = H + \sum_{\bf x} V(r) \, n_{\bf x}. \label{bhmt}
\end{eqnarray}
Therefore, the external trapping potential coupled to the particle
density turns out to be equivalent to an effective spatially-dependent
chemical potential
\begin{equation}
\mu_e({\bf x}) \equiv  \mu - V(r).
\label{mue}
\end{equation}
Far from the origin the potential $V(r)$ diverges, therefore the
particle density $\rho({\bf x})\equiv \langle n_{\bf x}\rangle$
vanishes and the particles are trapped.  We define the trap size by
\begin{equation}
l_t \equiv \sqrt{t}/v. 
\label{trside}
\end{equation}
This definition naturally arises~\cite{DSMS-96,BDZ-08,CV-10b} when we
consider the {\em thermodynamic} limit in a trap, which is generally
defined by the limit $N,l_t\to\infty$ keeping $N/l_t^3$ fixed ($N$ is
the number of particles). The value of the ratio $N/l_t^3$ is
controlled by the chemical potential $\mu$: it remains constant when
varying $l_t$ keeping $\mu$ fixed.

In the following, we set the hopping parameter $t=1$, so that all
energies are expressed in units of $t$, and the Planck constant
$\hslash=1$.

The inhomogeneity due to the trapping potential drastically changes,
even qualitatively, the general features of the behavior of homogenous
systems at a phase transition.  For example, the correlation functions
of the critical modes do not develop a diverging length scale in a
finite trap.  However, when the trap gets large the system shows a
critical regime, although distorted by the presence of the trap.
Around the transition point, i.e. when $T$ and $\mu$ are such that
$T\approx T_c(\mu)$, the correlation functions around the center of
the trap show power-law trap-size scaling behaviors with respect to
the trap size $l_t$, controlled by the XY universality class of the
critical behavior of the phase transition of the homogenous
system~\cite{CV-09}. The scaling behaviors of trapped BH models at the
BEC transition are discussed in Refs.~\cite{CTV-13,CN-14}, where also
numerical analyses are reported.

\section{Scaling behaviors of the  BEC}
\label{critspabec}

\subsection{Coexisting phases in the presence of the trap}
\label{phsp}

\begin{figure}
\includegraphics[width=7.5cm]{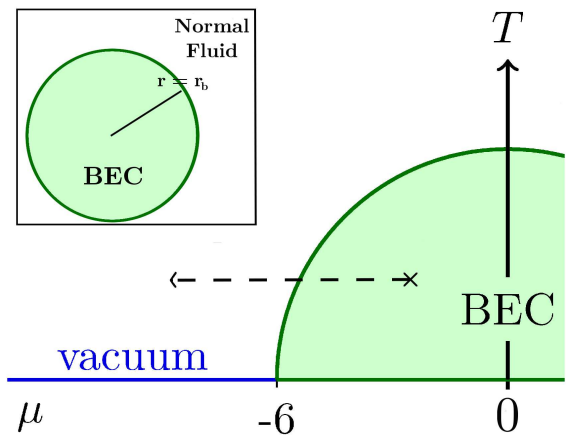}
\caption{We sketch the variation of the effective chemical potential
  $\mu_e(X) = \mu - X^2$ within the hard-core BH phase diagram, when
  $\mu<0$ and $0<T<T_c(\mu)$, indicated by the dashed line starting
  from $\mu_e(0)=\mu<0$.  The spatial BEC-to-normal transition occurs
  when $\mu_e(X)$ crosses the phase-boundary value $X_b\equiv
  r_b/l_t$.  The inset shows a section of the rotationally-symmetric
  system, with the corresponding phase regions: the BEC phase around
  the center of the trap and the normal phase for $r>r_b$.  }
\label{trapphases}
\end{figure}

The inhomogeneous conditions due to the harmonic trapping potential,
such as the BH model in Eq.~(\ref{bhmt}), may give rise to separate
regions where the quantum gas appears in the different BEC and normal
phases.

For example, let us again consider the phase diagram sketched in
Fig.~\ref{3dphasedia}, with the phase boundary $T_c(\mu)$.  When
$T<T_c(\mu)$ the quantum gas at the center of the trap is effectively
in the BEC phase.  Moving from the center, the effective chemical
potential
\begin{equation}
\mu_e({\bf x}) \equiv \mu_e(\mu;X)=\mu- X^2,\qquad X=r/l_t,
\label{muedep}
\end{equation} 
decreases. For sufficiently smooth changes, i.e. large trap size
$l_t$, we can apply the local density approximation (LDA): the local
observables around the spatial coordinate ${\bf x}$, and in particular
the particle density, can be approximated by those of the homogeneous
system at the same temperature and effective chemical potential
$\mu_e({\bf x})$.  When $r\equiv |{\bf x}|$ increases, $\mu_e = \mu -
X^2$ decreases, until it reaches a region for which
$T_c[\mu_e(\mu;X)]<T$, see Fig.~\ref{trapphases}.  Thus, we pass from
the BEC phase around the center of the trap to a region where the
quantum gas is in the disordered normal phase.  This scenario is
sketched in Fig.~\ref{trapphases}.  Analogous arguments apply to
soft-core BH models with a finite on-site interaction parameter $U$.

\subsection{Scaling behavior around the center of the trap}
\label{scbhc}

The space-coherence properties of the quantum gas within the BEC phase
are essentially associated with spin-wave modes. For $T\ll T_c$, when
the density $n_0$ of the condensate is much larger than the density of
the noncondensed atoms, the particle-field operator of homogenous
systems can be approximated~\cite{PSW-01} by $b({\bf x}) \approx
\sqrt{n_0} e^{i\varphi({\bf x})}$.  Then, the long-distance modes of
the phase correlations are expected to be described by an effective
spin-wave theory for the phase field $\varphi({\bf x})$, which is
invariant under a global shift $\varphi({\bf x})\to \varphi({\bf x}) +
\alpha$. The simplest spin-wave Hamiltonian is given by $H_{\rm sw} =
\int d^3x\; (\nabla \varphi)^2$.  Actually, the region where the
spin-wave theory effectively describes the long-distance phase
correlations extends to the whole BEC phase~\cite{PSW-01,CDMV-16},
i.e. for $T\lesssim T_c$, excluding only the relatively small critical
region close to $T_c$.  Therefore, the two-point spin-wave function
\begin{equation}
G_{\rm sw}({\bf x-y}) = \langle e^{-i\varphi({\bf x})} \,
e^{i\varphi({\bf y})} \rangle
\label{twopsw}
\end{equation}
is expected to describe the long-range phase-coherence properties of
homogenous particle systems in the whole BEC phase. Computations of
$G_{\rm sw}$ for various shapes and boundary conditions have been
reported in Ref.~\cite{CDMV-16}.

In the spin-wave limit, the inhomogeneity arising from the trapping
potential $V(r)$ may be taken into account by considering the
spin-wave Hamiltonian~\cite{CV-11}
\begin{equation}
H_{\rm sw} = \int d^3 x \, [1 + V(r)] (\nabla \varphi)^2,
\label{swham}
\end{equation}
where the potential modulates the coefficient related to the
superfluid density.  Due to the Gaussian nature of the spin-wave
theory, the scaling behavior of the correlation functions can be
inferred by a straightforward dimensional analysis~\cite{CV-11}.
Therefore, the relevant spatial scaling variable is expected to be the
ratio $X\equiv r/l_t$.  For example, if we consider the one-particle
correlation function, cf. Eq.~(\ref{gbdef}), between the center of the
trap and any point around the center, we expect the scaling behavior
\begin{equation}
G_0({\bf x};l_t) \equiv G({\bf 0},{\bf x};l_t) \approx g_0(X), 
\qquad X=r/l_t.
\label{g0x}
\end{equation}
These considerations can be straightforwardly extended to correlations
between generic points around the center, and to other observables.

\subsection{Scaling behavior at  the boundary of the BEC}
\label{scbhb}

We now argue that, for smooth trapping potentials, the quantum gas
develops a peculiar critical behavior around the spatial surface
separating the BEC and normal-phase regions, where
\begin{equation}
T_c[\mu_e(\mu;X)] \approx T < T_c(\mu).
\label{tcmue}
\end{equation}
For example, consider the hard-core BH lattice gas (\ref{bhmt}) for
$\mu<0$ and $T<T_c(\mu)$, see Fig.~\ref{trapphases}.  Since $T_c(\mu)$
decreases with decreasing $\mu$, a spherical surface exists at
distance $r_b$ such that
\begin{eqnarray}
&&T_c[\mu_e(\mu;X_b)] = T,
\label{tbou}\\
&&X_b \equiv  r_b/l_t= \sqrt{\mu-\bar{\mu}},
\qquad T_c(\bar{\mu})=T.
\label{xb}
\end{eqnarray}
This surface separates the BEC region from the normal-fluid region.
Expanding the rotationally-invariant potential $V(r)$ around $r_b$, we
obtain
\begin{equation} 
V(r) =V(r_b) + {2 X_b z/l_t} + {z^2/l_t^2},\qquad
z \equiv r - r_b.
\label{expvr}
\end{equation}
Therefore, the quantum gas around the surface at distance $r_b$
effectively behaves as a BH model with chemical potential $\bar{\mu} =
\mu_e(\mu;X_b)$ such that $T=T_c(\bar{\mu})$, in the presence of an
effectively linear potential $V_l(r)$ along the radial direction,
given by
\begin{eqnarray}
V_l(z) =   {z \over l_b},
\qquad l_b = {l_t \over 2 X_b}.
\label{vedef}
\end{eqnarray}
Since $X_b(\mu,T,l_t)>0$ is assumed finite and fixed, $l_b\sim l_t$.

We now derive some scaling ansatz meant to describe the asymptotic
behavior of the observables around the critical surface, in the limit
of large $l_b$. For this purpose, we note that the relevant spatial
variable is expected to be the distance from the critical surface at
$r=r_b$, i.e. $z = r-r_b$. This value should be compared with the
length scale $\xi_b$ of the critical correlation functions around the
surface $r=r_b$. Like general critical phenomena, see, e.g.,
Ref.~\cite{PV-02}, the asymptotic scaling behavior of the length scale
is expected to be characterized by a power law:
\begin{equation}
\xi_b\sim l_b^\theta,
\label{xieltheta}
\end{equation}
where $\theta$ is an appropriate exponent.  Therefore, the relevant
scaling variable is expected to be given by the ratio
\begin{equation}
Y \equiv {z/l_b^\theta}.
\label{defY}
\end{equation}

According to Eq.~(\ref{xieltheta}), $\theta$ is the exponent that
controls the relation between the length scale $\xi_b$ around the
transition point $r_b$ and the length scale $l_b$ of the linear
variation of the potential around $r_b$.  The exponent $\theta$ can be
determined by a scaling analysis of the perturbation associated with
the external linear potential coupled to the particle density.
Details are reported in the appendix.  The exponent $\theta$ turns out
to be related to the correlation-length exponent $\nu$ of the
universality class of the critical behavior of the homogeneous BEC
transition, i.e.,
\begin{equation}
\theta  = {\nu\over 1 + \nu} = 0.40181(3),
\label{theta}
\end{equation}
where $\nu=0.6717(1)$ is the correlation-length exponent of the 3D XY
universality class.

We may apply the above considerations to the correlation function
$G_r(r_1,r_2)$ along an arbitrary radial direction, i.e. along the
points with spherical coordinates ${\bf x}=(r,\theta,\varphi)$ keeping
the angle $\theta$, $\varphi$ fixed. We put forward the scaling
behavior
\begin{equation}
G_r(r_1,r_2;l_t,T) \approx 
\xi_b^{-1-\eta} {\cal G}_r(Y_1,Y_2),
\label{grsca}
\end{equation} 
where $Y_i=z_i/l_b^\theta$ with $z_i \equiv r_i - r_b$, and $r_1\neq
r_2$ (for equal points the scaling behavior differs, since
$G_r(r,r)=\rho(r)$, see below).  The prefactor is analogous to that
reported in Eq.~(\ref{hocrbeh}) for the critical behavior of
homogenous systems.  The difference among the different radial
directions, due to the lattice structure of the BH model, is expected
to be suppressed in the large-$l_t$ limit. Analogous scaling relations
can be straightforwardly derived for correlations between generic
points around the boundary between the two phases.

A scaling ansatz can be analogously derived for the connected
density-density correlation function
\begin{equation}
A({\bf x},{\bf y}) \equiv 
\langle n_{\bf x} n_{\bf y} \rangle - \langle n_{\bf x} \rangle 
\langle n_{\bf y} \rangle.
\label{ddcorr}
\end{equation}
Along the radial direction,
\begin{equation}
A_r(r_1,r_2;l_t,T) \approx 
\xi_b^{-2y_n} {\cal A}_r(Y_1,Y_2), \qquad r_1\neq r_2,
\label{crsca}
\end{equation} 
where $y_n=3-1/\nu$ is the renormalization-group dimension of the
density operator at the XY fixed point~\cite{CTV-13}.

The above scaling results can be straightforwardly extended to
different geometries of the trap and generic space dependences of the
external potential, when BEC and normal phases appear simultaneously
due to the inhomogeneous conditions.  Indeed, the linear approximation
at the spatial surface between the two phases is quite general. It
arises from general spatial dependences of the potentials, because a
linear term generally appears in the expansion of the external
potential around the critical surface.  Using scaling arguments one
can show that the quadratic (and in general any higher-order) term of
the expansion of the potential, cf. Eq.~(\ref{expvr}), gives rise to
suppressed contributions with respect to the asymptotic behavior
(\ref{grsca}). Indeed, close to the BEC boundary, when considering the
scaling behavior at fixed ratio $Y\equiv z/l_b^\theta$ as in
Eq.~(\ref{grsca}), the asymptotic behavior is determined by the
smallest $O(z)$ power law, while the quadratic $O(z^2)$ term gives
only rise to $O(l_b^{-1+\theta})$ corrections to the leading behavior
(\ref{grsca}). Therefore, the precise form of the trapping potential
turns out to be irrelevant for our considerations on the critical
behavior around the spatial separation of the different phases.

Actually we expect that there are other power-law corrections to the
asymptotic behavior, i.e., those related to the critical behavior of
the corresponding homogenous system, and in particular with the
scaling-correction exponent $\omega$ of the 3D XY universality
class~\cite{CHPV-06,GZ-98,PV-02}, i.e. $\omega=0.785(20)$.  They also
induce corrections in the presence of a linear external potential.
They should be $O(\xi_b^{-\omega})$, thus
$O(l_b^{-\omega\theta})$. Since $\omega \theta = 0.315(8)$ is smaller
than $1-\theta=0.59819(3)$, the scaling corrections arising from the
leading irrelevant perturbation of the XY universality class provide
the leading asymptotic corrections to the problem at hand.

The particle density $\rho({\bf x}) \equiv \langle b_{\bf x}^\dagger
b_{\bf x}\rangle$ is expected to vary smoothly across the transition
surface.  Its $l_t\to\infty$ asymptotic behavior approaches a finite
value, even at the transition surface.  This should be given by its
LDA, i.e., by the particle density $\rho_h$ of the homogenous system
at the corresponding values of the chemical potential and temperature,
i.e.,
\begin{equation}
\rho({\bf x}) \approx \rho_{\rm lda}({\bf x}) =
\rho_h[\mu_{e}(\mu;X),T].
\label{rhox}
\end{equation}
Thus the $l_t\to\infty$ limit of the particle density is expected to
be a function of the ratio $X\equiv r/l_t$.  Analogously to the
particle density of homogenous systems at criticality~\cite{CTV-13},
the critical behavior at the boundary of the BEC region is expected to
give rise to a nonanalytic subleading
$O(\xi_b^{-y_n})=O(l_b^{-y_n\theta})$ contribution, where
\begin{equation}
y_n\theta = {3\nu-1\over \nu+1} \approx 0.607.
\label{ynth}
\end{equation}
An analogous behavior has been also put forward, and numerically
checked, in the case of the 2D Ising model in the presence of a
temperature gradient~\cite{BDV-14}.

We also note that more complicated scenarios may appear in the
presence of inhomogeneous trap. For example, assuming the phase
diagram of Fig.~\ref{3dphasedia} for the hard-core $U\to\infty$ limit,
when $T<T_c(\mu=0)$, if $\mu>0$ is larger than the critical value
$\mu_c(T)>0$ (inverse function of $T_c(\mu)$ for $\mu>0$), then the
gas at the center of the trap is in the disordered normal phase. At
the distance such that $\mu_{e}(r)=\mu_c(T)$ it passes to a superfluid
phase until $\mu_e(r) = -\mu_c(T)$, where it passes again to a normal
phase. In this case we have two transition surfaces.  The critical
behaviors at such surfaces, in the regime of smooth external
potential, is expected to show scaling behaviors analogous to
Eq.~(\ref{grsca}). The exponent $\theta$ is the same of that given in
Eq.~(\ref{theta}), because at these transition surfaces the spatial
variation of the external potential is effectively linear.
 
Finally, we mention that quantum zero-temperature transitions between
spatially separated phases have been discussed in
Refs.~\cite{CV-10b,CTV-12} for low-dimensional quantum gases, between
the superfluid and vacuum phases.  They present similar scaling
behaviors, although the quantum nature of their critical modes makes
these phenomena substantially different from those at finite
temperature, as those discussed in this paper.

\section{Numerical results}
\label{numres}

In order to check the scaling behaviors put forward in the previous
section, we consider the 3D BH model in the hard-core $U\to\infty$
limit, whose phase diagram is sketched in Fig.~\ref{3dphasedia}.  We
present numerical results for the hard-core BH model at $\mu = -3$ and
$T = 1.482$ in the presence of a harmonic trap, cf. Eq.~(\ref{bhmt}).
The chosen value of the temperature corresponds to the critical
temperature for $\mu = -4$, indeed~\cite{CTV-13}
$T_c(\mu=-4)=1.4820(2)$.  This choice allows us to determine the
critical surface with great accuracy, see below, avoiding the
numerical uncertainty related to its location.  Of course, whenever
$T<T_c(\mu)$, the main features of the BEC scaling behaviors of
trapped BH models are expected to be largely
independent of this particular choice of parameters.

The trap is centered in the middle of a cubic $M^3$ lattice with open
boundary conditions. We consider odd sizes $M=2L+1$, so that the
lattice-site coordinates are $x_i=-L,...,L$, and the trap center is at
${\bf x}={\bf 0}$.  Numerical results are obtained by quantum Monte
Carlo (QMC) simulations using the directed operator-loop algorithm
\cite{SK-91,SS-02,DT-01} (see Ref.~\cite{CTV-12} for further details).
We report results for several values of the trap size $l_t$ (up to
$l_t=12$), for sufficiently large lattice size $L$ (up to $L=22$) to
obtain the infinite-$L$ limit of the quantities we are interested in,
within the typical statistical errors of our computations (this is
easily checked by comparing data for increasing values of $L$, see
also below).

\begin{figure}
\includegraphics[width=9cm]{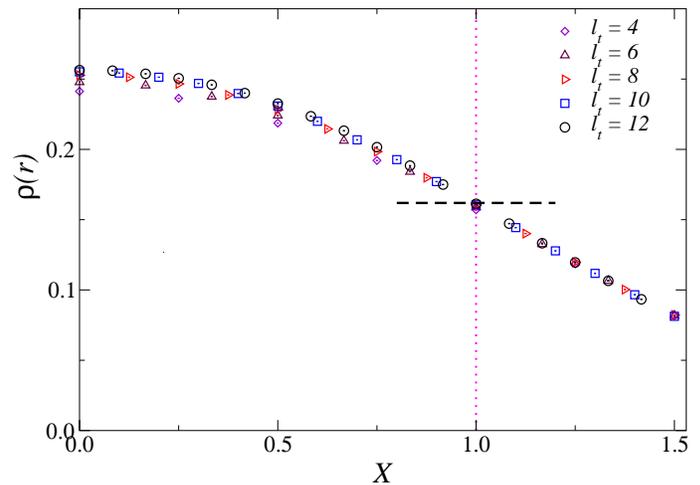}
\caption{Profile of the particle density versus $X\equiv r/l_t$, for
  $T=1.482$, $\mu=-3$ and various values of $l_t$.  The data are taken
  for sufficiently large lattice size to make finite-size effects
  negligible, i.e. smaller than the typical statistical errors (error
  bars are reported but they are so small to be hardly visible).  With
  increasing $l_t$, they appear to collapse toward a function of $X$,
  as predicted by LDA, cf. Eq.~(\ref{rhox}).  The vertical dashed line
  indicates the location $X_b\equiv r_b/l_t = 1$ of the spatial
  BEC-to-normal phase transition.  The horizontal bar indicates the
  LDA of the particle density at the critical surface, i.e.
  $\rho_{\rm lda}(X_b)=0.16187(1)$.  }
\label{density}
\end{figure}

\begin{figure}
\includegraphics[width=9cm]{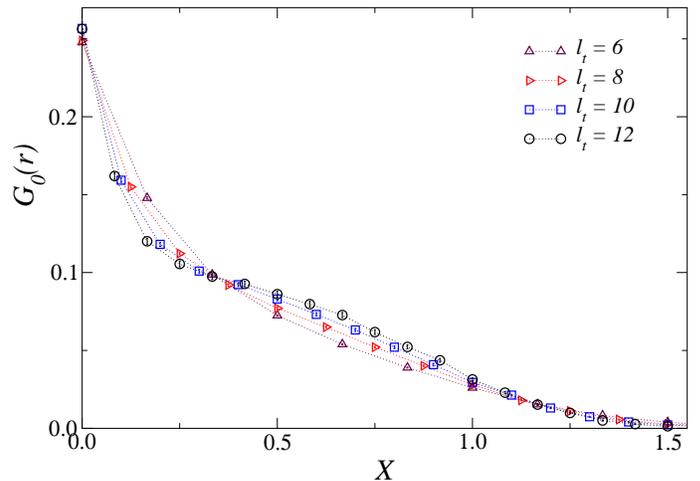}
\caption{Data for the one-particle correlation function $G_0(r) \equiv
  G({\bf 0},{\bf x})$ with $r\equiv |{\bf x}|$ versus $X\equiv r/l_t$,
  for $T=1.482$, $\mu=-3$ and some values of the trap size $l_t$.
  With increasing $l_t$, they appear to behave as a function of $X$,
  in agreement with the spin-wave predictions.  Note that $G_0(0) =
  \rho(0)$.}
\label{centercorrelator}
\end{figure}

According to the LDA discussed in the previous section, for
sufficiently large trap size $l_t$, the particle density $\rho({\bf
  x})\equiv \langle b_{\bf x}^\dagger b_{\bf x} \rangle$ is expected
to be a function of the local effective chemical potential $\mu_e({\bf
  x}) = \mu - X^2$ where $X\equiv r/l_t$, thus a function of $X$. This
is confirmed by the data for the density profile shown in
Fig.~\ref{density}. They clearly approach a function of $X$ with
increasing $l_t$. We expect that their convergence is generally
characterized by $O(l_t^{-1})$ corrections (related to the effective
linear variation of the potential around points at $r>0$), except at
the boundary of the BEC region, where it is expected to be slower (due
to the critical modes, see below), and at center of the trap where the
convergence is expected to be $O(l_t^{-2})$ (here the spatial
dependence of the potential is quadratic).  The numerical data turn
out to be consistent. For example, for $X=1/2$, where the effective
chemical potential has a linear slope, the data nicely fit $\rho_{1/2}
+ cl_t^{-1}$ (we obtain $\rho_{1/2}=0.2411(3)$ and $\chi^2/{\rm
  d.o.f}\approx 0.5$ using the data for $l_t\ge 6$).  At the center of
the trap, the data favor the behavior $\rho_0 + cl_t^{-2}$ (we obtain
$\rho_0=0.2592(4)$ and $\chi^2/{\rm d.o.f}\approx 0.1$ from the data
for $l_t\ge 6)$.

The particle density $\rho({\bf x})$ shows a smooth behavior also
across the transition surface at $r_b=l_t$, see Fig.~\ref{density}.
Indeed, like the energy density at a continuous
transition~\cite{PV-02}, its large-$l_b$ limit is expected to be a
nonuniversal constant given by its LDA, while critical modes give rise
to subleading $O(\xi_b^{-y_n})$, thus $O(l_b^{-y_n\theta})$,
contributions, cf. Eq.~(\ref{ynth}).  Therefore, we expect
\begin{equation}
\rho(r_b) =\rho_{\rm lda} + O(l_b^{-y_n\theta}),
\label{rhorb}
\end{equation}
where $\rho_{\rm lda}$ is the value of the particle density of the
homogenous system at $\mu=-4$ and $T_c(\mu=-4)=1.482$,
i.e.~\cite{CTV-13} $\rho_{\rm lda}=\rho_c(\mu=-4) = 0.16187(1)$.  The
data for the particle density $\rho(r_b)$ are consistent, see
Fig.~\ref{density}; they approach $\rho_{\rm lda}$, indeed $\rho(r_b)
= 0.1600(2),\, 0.1608(2),\,0.1613(2)$ for $l_t=8,\,10,\,12$
respectively.

Figure~\ref{centercorrelator} shows data for the one-particle
correlation function $G_0(r)$ between the center and a generic point,
cf. Eq.~(\ref{g0x}).  Their behavior is substantially consistent with
the large-$l_t$ scaling behavior (\ref{g0x}) inferred by
spin-wave arguments, i.e., $G_0(r)\approx g _0(X)$ in the large
trap-size limit (with a $X$-dependent convergence rate).

We now consider the behavior at the spatial boundary of the BEC, where
the quantum gas is expected to develop a critical behavior.  According
to the scenario put forward in Sec.~\ref{phsp}, the spatial boundary
between the BEC and normal phase regions is located at the surface
$r=r_b$ corresponding to the effective chemical potential
$\mu_e(\mu;X_b)$ were $X_b=r_b/l_t$, such that
$T=T_c[\mu_e(\mu;X_b)]$.  For our choice of parameters,
$\mu_e(\mu;X_b) = -4$, thus $r_b=l_t$.  Therefore, the effective
potential (\ref{vedef}) around the critical surface is characterized
by the length scale $l_b=l_t/2$.

\begin{figure}
\includegraphics[width=9cm]{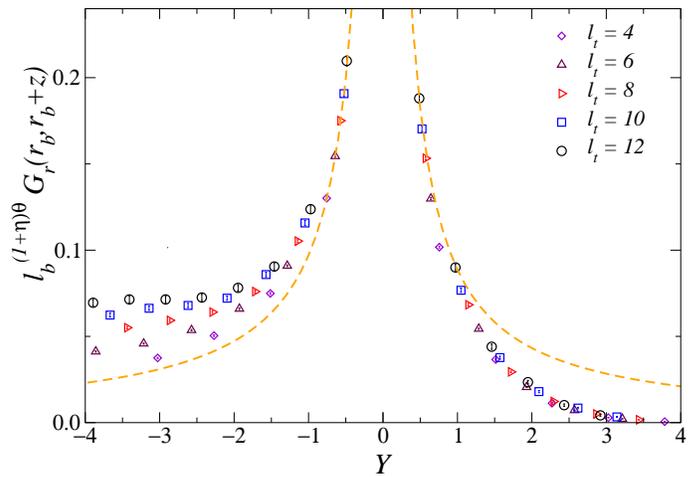}
\caption{ Scaling behavior of the one-particle correlation function
  $G_r(r_b,r_b+z)$ along a radial direction around the critical
  surface. We show data for $l_b^{(1+\eta)\theta}G_r(r_b,r_b+z)$ for
  $T=1.482$, $\mu=-3$ and some values of $l_t=2l_b$, versus $Y\equiv
  z/l_b^\theta$.  They are taken for sufficiently large lattice size
  to make finite-size effects negligible within errors.  With
  increasing $l_b$, the data appear to converge to a scaling function
  of $Y$, in agreement with Eq.~(\ref{grsca}). Note that positive
  (negative) values of $Y$ refer to the normal (BEC) phase regions.
  The dashed lines show the power law $c |Y|^{-1-\eta}$ expected for
  small values of $|Y|$.  }
\label{criticalcorrelator}
\end{figure}

In order to check the scaling ansatz (\ref{grsca}), we compute the
one-particle correlation function $G_r(r_b,r_b+z)$ along one of the
main axis of the lattice, around the critical point $r=r_b=l_t$.  The
data are shown in Fig.~\ref{criticalcorrelator}.  They nicely support
the scaling ansatz (\ref{grsca}). Indeed the data for the product
$l_b^{(1+\eta)\theta} G_r(r_b,r_b+z)$ appears to approach a scaling
function ${\cal G}_r(0,Y)$ of $Y\equiv z/l_b^\theta$ with increasing
$l_b$ [except at $z=0$ where such a scaling is not expected, because
  $G_r(r_b,r_b) = \rho(r_b)$, see Eq.~(\ref{rhorb})].  Note that the
data for small values of $|Y|$ are substantially consistent with the
asymptotic power-law behavior
\begin{equation}
{\cal G}_r(0,Y)\sim|Y|^{-1-\eta},
\label{matchg}
\end{equation}
matching the small-distance critical behavior (\ref{hocrbehtc}) of
homogenous systems. Therefore, in the neighborhood of the critical
surface, the one-particle correlation function shows the typical
singular behavior of critical phenomena, cf. Eq.~(\ref{hocrbehtc}),
i.e. when $|z|\ll l_b^\theta$
\begin{equation}
G_r(r_b,r_b+z)\sim |z|^{-1-\eta}.
\label{grb}
\end{equation}

In Fig.~\ref{criticalcorrelator} the corrections to the asymptotic
scaling function ${\cal G}_r(0,Y)$ appear larger for $Y<0$,
corresponding to the BEC phase region, while the scaling approach
turns out to be quite rapid for $Y>0$ corresponding to the normal
phase region.  As argued in the previous section, the leading
corrections to the asymptotic behavior are expected to be
$O(l_b^{-\omega\theta})$ due to the leading irrelevant perturbation of
the 3D XY universality class.  The effects of the higher-order terms
(beyond the linear one) in the expansion of the potential around the
critical surface, cf. Eq.~(\ref{expvr}), are expected to be more
suppressed, i.e., they are $O(l_b^{-1+\theta})$.

Analogous results are expected along the other directions, indeed the
effects of the breaking of the rotational symmetry due to the lattice
structure are expected to be $O(l_t^{-1})$
suppressed in the large-$l_t$ limit.

\begin{figure}
\includegraphics[width=9cm]{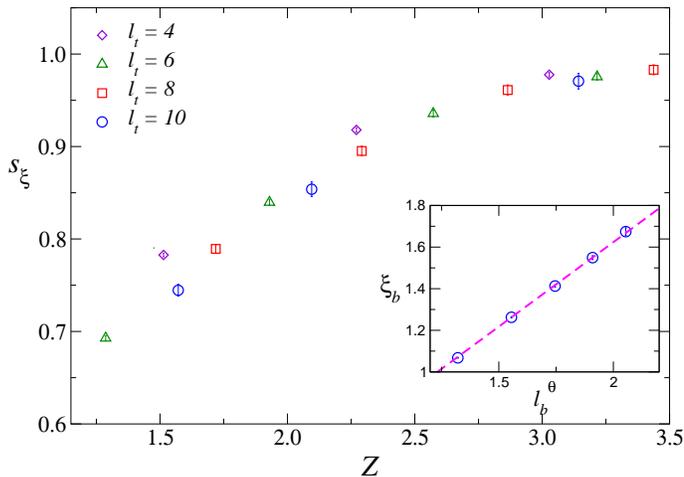}
\caption{QMC data for the ratio $s_\xi(L) =
  \xi_b(L)/\xi_b(L\to\infty)$ with $\xi_b$ defined as in
  Eq.~(\ref{xibdef}), versus $Z\equiv (L-r_b)/l_b^\theta$.  With
  increasing $l_t$, the data appear to approach a scaling function of
  $Z$.  The inset shows data of $\xi_b$ for the largest available
  lattices, corresponding to $Z\gtrsim 4.5$, which turn out to be
  sufficient to provide an accurate estimate of $\xi_b(L\to\infty)$
  (we use these data to estimate $\xi_b(L\to\infty)$ in the ratio
  $s_\xi$).  They clearly show the predicted scaling behavior
  $\xi_b(L\to\infty)\sim l_b^\theta$; the dashed line shows the fit
  to $\xi_b=c\,l_b^\theta$ with $\theta$ given by Eq.~(\ref{theta}).
      }
\label{fss}
\end{figure}

We finally discuss the finite-size effects arising from the finite size
$M=2L+1$ of the lattice.  The above results have been obtained for
sufficiently large $L$ to make finite-size effects negligible.  This
condition can be made more precise.  Since the length scale of the
critical modes around the critical sphere is $\xi_b\sim l_b^\theta$,
we expect that finite-size effects get suppressed when $(L-r_b)/\xi_b
\gg 1$, thus $(L-r_b)/l_b^\theta \gg 1$, where $L-r_b$ is the distance
of the critical sphere from the boundary of the lattice.  Actually, we
expect that the interplay between $L$ and $l_b$ leads to finite-size
scaling (FSS) behaviors analogous to FSS in homogenous
models~\cite{Cardy-88}.  In the case at hand the relevant scaling
variable is expected to be given by the ratio
\begin{equation}
Z \equiv (L-r_b)/l_b^\theta .
\label{Zdef}
\end{equation}

In order to show this further feature of the critical
behavior at the boundary of the superfluid region, 
we consider the correlation length $\xi_b$ of the critical
modes around the critical surface, defined as
\begin{equation}
\xi_b(L)^2 = {\sum_{z=0}^{L-r_b} z^2 G_r(r_b,r_b+z) \over
\sum_{z=0}^{L-r_b} G_r(r_b,r_b+z) }
\label{xibdef}
\end{equation}
where the sum is meant along one of the main directions of the
lattice.  Eq.~(\ref{grsca}) implies that
\begin{equation}
\xi_b \equiv \xi_b(L\to\infty)\sim l_b^\theta. 
\label{xibsca}
\end{equation}
This is confirmed by the data of $\xi_b$ shown in the inset of
Fig.~\ref{fss}, obtained for sufficiently large lattice sizes
$L>r_b=l_t$ to provide a good approximation of $\xi_b(L\to\infty)$
(within an accuracy of one per cent, as checked by increasing $L$ at
fixed $l_t$).  Indeed, the data from $l_b=2$ to $l_b=6$ nicely fit the
ansatz $\xi_b = c\,l_b^\theta$ with $\theta$ given by
Eq.~(\ref{theta}), i.e. $\theta=0.40181(3)$, see the inset of
Fig.~\ref{fss}. In particular, considering $\theta$ as a free
parameter, we obtain a consistent value with an accuracy of a few per
cent, i.e. $\theta=0.407(3)$ using all available data, with an
acceptable $\chi^2/{\rm d.o.f.}\approx 0.1$.

Then, we compute the ratio
\begin{equation}
s_\xi(L) = \xi_b(L)/\xi_b(L\to\infty),
\label{sxi}
\end{equation}
for various values of $l_t$ and $L$.  The FSS hypothesis implies that
$s_\xi$ must approach a scaling function of $Z$,
cf. Eq.~(\ref{Zdef}). This is nicely supported by the data reported in
Fig.~\ref{fss}.  This FSS analysis shows that the condition $Z \gtrsim
4$ is sufficient to provide the $L\to\infty$ limit of $\xi_b$ within
approximately one per cent.

We finally note that the problem that we consider, i.e. the critical
behavior at the boundary of the BEC region, differs from the phenomena
of boundary critical behaviors at the boundaries of statistical
systems~\cite{Binder-83,Diehl-86}.  In the latter case the
inhomogeneous behavior arises from the presence of the boundaries. In
the case that we study the inhomogeneous critical behavior is
controlled by the external effectively linear potential, giving rise
to a critical region of size $\xi\sim l_b^\theta$ separating the BEC
and normal-phase regions.

\section{Conclusions}
\label{conclu}

We investigate some aspects of the BEC of quantum gases in the
presence of inhomogeneous conditions, such as cold-atom systems in
harmonic traps, which are typically realized in experiments, see,
e.g.,
Refs.~\cite{Andrews-etal-97,Stenger-etal-99,Hagley-etal-99,BHE-00,
  Cacciapuoti-etal-03,Hellweg-etal-03,Ritter-etal-07,BDZ-08,DGPS-99,
  PSW-01,Dettmer-etal-01,Hellweg-etal-02}.  We consider 3D bosonic
gases trapped by an external potential, when the temperature is
sufficiently low to show a superfluid phase region around the center
of the trap.  In particular, we consider 3D BH models in the presence
of a harmonic trap, cf. Eq.~(\ref{bhmt}), which model realistic gases
of bosonic atoms in optical lattices~\cite{JBCGZ-98}.

We point out that, if the trap is sufficiently large, different phases
may coexist in different space regions, when moving from the center of
the trap. For example, we may pass from the superfluid BEC phase
around the center of the trap (where space coherence is essentially
described by spin waves) to a normal phase far from the center.
Between the superfluid and normal phase regions the quantum gas
experiences a {\em spatial} phase transition.  The system develops a
peculiar critical behavior at the surface separating the different
phases, characterized by the presence of an external potential with an
effectively linear space dependence. Using scaling arguments, we put
forward the asymptotic behavior in the limit of smooth external
potential, i.e., large length scale $l_b$ associated with the
effective linear variation of the potential at the critical surface,
cf. Eq.~(\ref{vedef}).  We argue that this peculiar scaling behavior
is controlled by the universality class of the homogenous BEC
transition, i.e., the 3D XY universality class whose critical
exponents are known with great accuracy. The length scale $\xi_b$ of
the critical modes around the critical surface diverges as $\xi_b \sim
l_b^\theta$ with $\theta = \nu/(1 + \nu)$, where~\cite{CHPV-06}
$\nu=0.6717(1)$ is the correlation-length critical exponent of the 3D
XY universality class.

We provide numerical evidence of this phenomenon for the 3D BH model
in the hard-core $U\to\infty$ limit (whose phase diagram is sketched
in Fig.~\ref{3dphasedia}), in the presence of a rotationally-invariant
harmonic trap.  The numerical results, obtained by QMC simulations,
nicely support our scaling predictions, in particular those at the
boundary of the BEC.  An analogous scenario is expected for soft-core
BH models with finite on-site interaction parameters $U$.

The scaling theory at the boundary of the BEC can be straightforwardly
extended to different geometries of the trap, and/or generic space
dependences of the external potential.  This is essentially due to the
fact that the linear approximation at the spatial surface between the
two phases is quite general. Moreover, high-order terms are expected
to be irrelevant: they give only rise to $O(l_b^{-1+\theta})$
suppressed contribution in the $l_b\to\infty$ limit.

We remark that the lattice structure of the BH model does not play any
particular role in our scaling arguments concerning the spatial phase
transition at the boundary of the BEC, and the derivation of the
corresponding power laws. Indeed the microscopic details of the model
are irrelevant in the asymptotic $l_b\to\infty$ limit. Therefore, our
scaling predictions at the boundary of the BEC apply to a wide class
of models, i.e. to any 3D inhomogeneous interacting bosonic systems at
the spatial boundary of their BEC phase.

The experimental evidence of the critical behavior at the boundary of
the BEC in trapped quantum gases requires the measurement of the
one-particle correlation functions, or higher-order correlations.
These quantities are not easily accessible. However several examples
of such space-coherence measurements have been reported in the
literature, see, e.g.,
Refs.~\cite{Hagley-etal-99,BHE-00,DRBOKE-07,NGSH-15}.  On the other
hand, the more accessible particle density, which can be measured by
{\em in situ} density image techniques~\cite{GZHC-09,BGPFG-09}, can
hardly provide evidence of this critical phenomenon, since the
critical modes give only rise to subleading contributions to the
particle density, cf. Eq.~(\ref{rhorb}).

\appendix

\section{Derivation of the exponent $\theta$ }
\label{thetaest}

The exponent $\theta$ can be inferred by a scaling analysis of the
perturbation $P_V$ representing the external linear potential
$V_l(r)=u r$ coupled to the particle density.  We follow the
field-theoretical approach of Refs.~\cite{CV-09,CV-10}, that is we
consider the 3D $\Phi^4$ quantum field theory which represents the 3D
XY universality class, see e.g.  Ref.~\cite{ZJ-book},
\begin{equation}
H_{\Phi^4} = \int d^3 x                                        
\left[ |\partial_\mu \psi({\bf x})|^2 + 
r |\psi({\bf x})|^2 + u |\psi({\bf x})|^4\right],
\label{hphi4}
\end{equation} 
where $\psi$ is the complex field associated with the order parameter,
and $r,u$ are coupling constants.  Since the particle density
corresponds to the energy operator $|\psi|^2$, we can write the
perturbation $P_V$ as
\begin{equation}
P_V=\int d^3 x\, V_l({\bf x}) |\psi({\bf x})|^2.
\label{pertu}
\end{equation}
Introducing the renormalization-group dimension $y_u$ of the constant
$u$ of the linear potential, we derive the scaling relation $y_u - 1 +
y_n = 3$, where $y_n=3-1/\nu$ is the RG dimension of the
density/energy operator $|\psi|^2$ (we recall that $\nu$ is the
length-scale critical exponent of the 3D XY universality class). We
eventually obtain $\theta\equiv 1/y_u$, and therefore
Eq.~(\ref{theta}).

\end{document}